**Rapid Production of Accurate Embedded-Atom Method Potentials for Metal Alloys**

Logan Ward[1], Anupriya Agrawal[1], Katharine M. Flores[2], and Wolfgang Windl[1]

[1] Department of Materials Science and Engineering, The Ohio State University

[2] Department of Mechanical Engineering and Materials Science, Washington University

Abstract

The most critical limitation to the wide-scale use of classical molecular dynamics for alloy design is the availability of suitable interatomic potentials. In this work, we demonstrate a simple procedure to generate a library of accurate binary potentials using already-existing single-element potentials that can be easily combined to form multi-component alloy potentials. For the Al-Ni, Cu-Au, and Cu-Al-Zr systems, we show that this method produces results comparable in accuracy to alloy potentials where all parts have been fitted simultaneously, without the additional computational expense. Furthermore, we demonstrate applicability to both crystalline and amorphous phases.

Body

1. Introduction

Classical molecular dynamics (MD) simulations have become ubiquitous tools for simulating complex behavior in many materials systems from atomistic to increasingly larger length scales. One of the key issues limiting the ability for materials scientists to design metallic materials from the atom up using classical MD is the lack of interatomic potentials suitable for the desired alloy system. Interatomic potentials, which approximate complex interatomic interactions using computationally efficient functions, range from simple pair interactions to more complex formulations that involve the local electron density and bond angles. Of these, the embedded-atom method (EAM) has been used widely because of its low computational costs and ability to accurately model bulk properties and defects in metals[1,2].

The EAM is a many-body interatomic potential consisting of a pair function and a many-body interaction term. In the Finnis-Sinclair form of the embedded atom method[2], the energy of a single atom is computed as

$$U_i = F_\alpha\left(\sum_{i\neq j} f_{\alpha\beta}(r_{ij})\right) + \frac{1}{2}\sum_{i\neq j}\phi_{\alpha\beta}(r_{ij}) \qquad (1)$$

where the sum is over all atoms less than a cutoff distance apart. The key features of this formula are a pair-interaction term, $\phi$, and an embedding function, $F$, that depend non-linearly on the contributions from the neighboring atoms to the local electron density, $f$. In this form, the pair-interaction and electron density functions are different for each combination of central atom type ($\alpha$) and neighbor type ($\beta$), whereas the embedding functions are specific to the species of each neighbor, $\beta$, and central atom, $\alpha$, respectively.

Whereas good-quality EAM potentials exist for a considerable number of elemental metals[3–5], potentials for alloy systems are sparse and typically require considerable development effort[6], which makes fast exploration of new alloy compositions rather difficult. Interatomic potentials describing alloys are often developed by coupling several previously-developed pure element potentials and fitting the cross-interaction functions against experimental or *ab initio* data from the alloy system (i.e. mixing enthalpies and moduli)[7,8]. Along with fitting the cross-interaction terms, it is possible to increase the quality of the potential by making alterations to the pure-species functions that effect the calculated alloy energies, but leave pure element energies unaffected[7]. These alterations to the elemental components are unique to each fitted binary system, which means they must be adjusted when adding additional elements to a potential. This makes the development of many-component alloy potentials slow and limits the accuracy as more components are added, as the same transformations must satisfy even more objectives. If one seeks to mitigate this problem by fitting pure and alloy functions simultaneously, the number of fitting parameters becomes impractical to optimize for systems with more than 4 or 5 components[6].

The Johnson alloy model was proposed with the goal of creating multi-component EAM potentials without significant computational expense[9]. The basic idea of this method is to generate the new pair functions for dissimilar species (cross-potentials) required to create an alloy potential as electron density-weighted averages of the elemental components based on an empirical model,

$$\phi_{\alpha\beta}(r) = \frac{1}{2}\left(\frac{f_\beta(r)}{f_\alpha(r)}\phi_{\alpha\alpha}(r) + \frac{f_\alpha(r)}{f_\beta(r)}\phi_{\beta\beta}(r)\right). \qquad (2)$$

It is important to note that this method uses a different version of the EAM in which the electron density function ($f_\alpha$) is only specific to a single-atom type rather than to a pair of types as in the Finnis-Sinclair EAM potential. In order to ensure the best-possible alloy potentials between many different elements, this method only requires finding compatible magnitudes for the electron density functions of all of the elements in a database and ensuring slope of the embedding function of each potential is zero at the equilibrium electron density for that species. The compatible magnitudes are set by introducing a scaling parameter for each electron density function that, in order to leave the elemental embedding energy unchanged, is compensated for by reciprocal scaling in the embedding function. The appropriate scaling factor can be determined by minimizing the total error in the dilute heats of solution for each possible binary in the database[9], or using empirical relationships based on the cohesive energy and atomic volume of each constituent[10,11]. In this work, we will demonstrate that this method, while fast, can significantly compromise the accuracy of interatomic interactions in alloy systems.

The method presented in this paper has computational expenses somewhat higher than those required for the Johnson alloy model, but is considerably more accurate than this model while also being faster than holistic fits encompassing the entire alloy system. We propose creating binary Finnis-Sinclair EAM potentials using high-quality elemental EAM potentials from the literature with only small modifications and cross-potentials components fit to a limited number of properties of a limited set of intermetallic phases calculated from first principles. We show that these binaries can be combined into many-component alloy potentials without significant alteration or additional fitting required, which opens a new path for large-scale exploration of multi-component alloys.

2. Method

The method proposed in this paper describes a generally applicable method to join elemental EAM potentials and fit the necessary alloy functions to data calculated using Density Functional Theory (DFT).

2.1. *Ab Initio* Calculations

In order to generate a training set, the equilibrium lattice parameter, mixing enthalpy, and bulk modulus are calculated using DFT for the B2 and both L1$_2$ intermetallics in the desired binary. In our implementation of this method, these are calculated using VASP with plane wave basis sets and Generalized Gradient Approximation (GGA) exchange-correlation[12–14]. The number of k-points for Brillouin zone integration used for each calculation is increased until the total energy changes by less than 0.01% with the addition of more k-points. Generally, this convergence is reached with a grid of 14×14×14 k-points.

In order to compensate for the difference between lattice parameters and bulk moduli predicted by DFT and measured experimentally, a simple "rule-of-mixtures" approach is used. For each element, correction factors for the lattice parameters and bulk modulus are determined using

$$C_{lattice} = \frac{1}{3}\left(\frac{a_{Expt.}}{a_{DFT}} + \frac{b_{Expt.}}{b_{DFT}} + \frac{c_{Expt.}}{c_{DFT}}\right) \tag{3}$$

$$C_{Modulus} = K_{Expt.}/K_{DFT} \tag{4}$$

The moduli and lattice parameters calculated for intermetallics are multiplied with an effective correction factor, which is an average of the elemental correction factors weighted by composition. Handbook values are easily available for the experimental data required for these correction factors[15–17].

### 2.2. Potential-Fitting Procedure

The following sub-section outlines methods to combine elemental potentials, fit cross-potential functions against *ab initio* data, and then join binaries into more-complex potentials.

#### 2.2.1. Standardizing Elemental Potentials

The single-element potentials are initially adapted to improve compatibility in binary and multicomponent potentials in a way that preserves the original accuracy in the pure systems. First, the embedding functions are adjusted to exist on the same range of arbitrary units and the electron density functions are scaled appropriately using the invariant transformations[7]

$$F(\rho) \rightarrow F\left(\rho/s\right), \tag{5}$$

$$f(r) \rightarrow s \cdot f(r). \tag{6}$$

where $\rho$ is an electron density, $s$ is an arbitrary scaling factor, and $r$ is a distance. Second, the pair interaction functions are adjusted to be of a qualitatively similar form. Depending on the functional form used in the original potentials, these functions can range from repulsive at all distances[18] to attractive at long distances[5]. In this work, we chose to transform the potentials into the latter form by adjusting the pair function such that its minimum is 4% of the cohesive energy using the invariant transformations demonstrated by Voter[7]

$$F(\rho) \rightarrow F(\rho) + g \cdot \rho \quad (7)$$

$$\phi(r) \rightarrow \phi(r) - 2g \cdot f(r) \quad (8)$$

where $g$ is the transfer parameter.

### 2.2.2. Joining Elemental Potentials

When joining two elemental potentials, the maximum cutoff distance between the two potentials is adopted for the binary potential. The electron density functions and pair interaction terms from the original potentials are defined as equal to 0 for distances greater than the original cutoff. As most interatomic potentials are available in a tabulated format with each function defined on discrete intervals, our implementation uses a cubic spline interpolation to evaluate each component functions on values not specified in the original table.

### 2.2.3. Fitting Cross-Potentials

The Finnis-Sinclair formulation for an EAM potential is used for each alloy potential. For a binary system, this requires two embedding functions ($F_\alpha, F_\beta$), four electron density functions ($f_{\alpha\alpha}, f_{\alpha\beta}, f_{\beta\alpha}, f_{\beta\beta}$), and three pair interaction functions ($\phi_{\alpha\alpha}, \phi_{\alpha\beta}, \phi_{\beta\beta}$). Since the functions involving only one atomic species are supplied by the single-element potentials, it is only necessary to fit three interspecies functions, which are two electron densities and one pair potential.

We chose to fit the electron density functions as scaled versions of the elemental functions, where

$$f_{\alpha\beta}(r) = s_\alpha \cdot f_{\alpha\alpha}(r), \quad (9)$$

$$f_{\beta\alpha}(r) = s_\beta \cdot f_{\beta\beta}(r). \quad (10)$$

This method adjusts the electron density function to compensate for the electron density cloud around an atom reacting differently depending on the adjacent species. In this form, we assume that the effective magnitude of the density clouds is different, but not the shape. The key advantage of this technique is that it only requires two fitting parameters, $s_\alpha$ and $s_\beta$, and does not require knowing the original functional forms of the electron density functions.

The pair interaction function is fitted using a Morse function, which is defined as

$$\phi_{\alpha\beta}(r) = E_1[e^{(-2\alpha(r-r_0))} - 2e^{(-\alpha(r-r_0))}]. \tag{11}$$

This function was chosen because its shape reflects the standardized pair potential shape specified in section 2.2.1, which is highly repulsive at short distances and slightly attractive at long distances. The Morse function was chosen in lieu of other functions with this shape because of its simplicity, with only three fitting parameters, $E_1$, $\alpha$, and $r_0$. Each parameter is allowed to vary over any range, with $\alpha > 0.9$ Å$^{-1}$ being the only constraint. This range was chosen based on the observation that $\alpha$ values smaller than approximately 0.9 Å$^{-1}$ lead to unphysical behavior in some of the tested intermetallic compounds.

In order to ensure that the function and its first derivative approach zero at the cutoff distance, the transformation

$$\phi_{\alpha\beta}(r)_{smooth} = \phi_{\alpha\beta}(r) - \phi_{\alpha\beta}(r_{cutoff}) + \left(\frac{r_{cutoff}}{m}\right)\left[1 - \left(\frac{r}{r_{cutoff}}\right)^m\right]\left(\frac{d\phi_{\alpha\beta}}{dr}\right)_{r=r_{cutoff}} \tag{12}$$

is applied to the pair function[19]. In all of the potentials generated in this study, $m$ was set to 20 as suggested in Ref. 19.

The optimum values of all five coefficients (two for the electron density and three from the pair function) are determined by minimizing the difference between the mixing enthalpy, bulk modulus, and lattice parameters of the B2 and L1$_2$ intermetallics calculated by our potential and corrected DFT (section 2.1). A genetic algorithm was chosen to optimize the objective function, which is the weighted sum of total error in mixing enthalpy, and fractional errors in bulk modulus and lattice parameter calculated using the relationship:

$$\frac{1}{F} = \sum_i \left[\frac{\Delta H_{m,i}}{w_H} + \left(w_K \cdot \frac{\Delta K_i}{K_{i,\text{DFT}}} + w_a \cdot \frac{\Delta a_i}{a_{i,\text{DFT}}}\right)\right]. \tag{13}$$

In this form, $F$ represents the fitness of the potential with respect to matching the corrected *ab initio* values. The weight factors $w_H$, $w_K$, and $w_a$ were set to be 0.4 eV, 1.0, and 1.6, respectively. The weights for each property were fixed by adjusting each parameter using a balanced approach until differences between the mixing enthalpies, bulk moduli and lattice parameters calculated using our potential and *ab initio* were equivalent to or smaller than those differences from an Al-Ni potential available in the literature[20], as described later. These weight factors were subsequently used for every binary potential created in our study.

*2.2.4.* Joining into many-component potentials

As with the single-component potentials, the binaries are combined by transforming the embedding functions to exist on the same range and adopting the maximum cutoff radius. As long as the same pure element potentials are used to create each binary, the electron density and pair interaction functions contained within the source binaries are sufficient to create higher-order potentials without any additional fitting. In addition, the properties of each binary system and the original pure elements are preserved in these new potentials.

3. Validation

Potentials created with the proposed technique were validated by comparison with experimental results and existing alloy potentials. In addition, the dependence of the accuracy of our binary potentials on the source of the elemental potentials was assessed.

3.1. Comparison to Experimental Values

Potentials were created for the Al-Ni and Cu-Au systems in order to evaluate their ability to replicate experimental data and to ensure DFT is suitable for providing a training set of intermetallic properties. For both systems, elemental potentials from Zhou *et al*. were used to generate interatomic potentials[5] and experimental data was taken from[21–26]. Elemental potentials from the same author were used to remove author compatibility as a factor for this test.

As shown in Table 1, there is generally good agreement of mixing enthalpy, bulk modulus, and lattice parameter between DFT, the fitted EAM potentials, and experiment. Also, the DFT correction factors to the bulk moduli and lattice parameters from Eq. (3) and (4) were found to be a necessary addition to this method. For example, the bulk modulus of $Au_3Cu$ calculated with DFT is 26 GPa (16%) below the experimental values without correction and only 1.4% after correction. Overall, the DFT values for mixing enthalpy and the corrected DFT values for bulk modulus and lattice parameter agree exceptionally well with experiment and are all within 0.075 eV, 10 GPa, and 0.03 Å of experimental values, respectively. In contrast, the uncorrected DFT values for bulk modulus and lattice constant deviated as much as 34 GPa and 0.09 Å. Based on this result, it was concluded that corrected DFT is suitable for creating a training set of intermetallic properties.

The degree that our potentials agree with corrected DFT was found to be strongly dependent on the weight factors used in Equation 13. As shown in Figure 1, it is possible to reproduce a single property extremely well at the expense of others by weighting that property heavily. Potentials that ignore key material properties are not desirable, so an even balance between them was adopted. To decide on the proper weighting between properties, the weight factors in Equation 13 were adjusted until differences between the mixing enthalpies, bulk moduli and lattice parameters calculated using our potential and *ab initio* values were equivalent to or smaller than those differences from an Al-Ni potential available in the literature[20]. The Al-Ni system was selected because of the availability of several Al-Ni potentials[5,20] and a large body of experimental data[21,24–26]. These weight factors were subsequently used for every binary potential created in our study.

The EAM potentials are able to replicate this training set with only small deviations from the corrected DFT data, as shown in Table 1. In the Al-Ni binary, the maximum error in mixing enthalpy, bulk modulus, and lattice parameter are very low at 0.015 eV, 10 GPa, and 0.07 Å, respectively. The maximum errors compare well to a recent potential developed to model NiAl and $Ni_3Al$ intermetallics[20], which has maximum errors of 0.17 eV, 29 GPa, and 0.06 Å. The maximum errors for our potential in the Cu-Au system are even lower at 0.009 eV, 7 GPa, and 0.04 Å, respectively. This Cu-Au compares favorably to a potential developed by Zhou *et al.* using the Johnson alloy model[5], which differs as much as 0.07 eV, 43.6 GPa, and 0.04 Å for the same three tested intermetallics. This shows that our method can generate binary potentials that match DFT and experimental properties just as well as, if not better than, other potentials available in literature.

### 3.2. Effect of Potentials from Different Authors

The Cu-Au binary was refit with elemental potentials for gold from other authors in order to assess the effect of using elemental potentials with a broad sample of potential fitting approaches[5,27–29]. For instance, the Ackland potential functions are based on fixed forms and fit to the Au lattice parameter, elastic constants, vacancy formation energy, cohesive energy, and the stacking fault energy[28]. In contrast, the Au potential developed by Sheng used quintic splines fit using the force-matching approach and a large database of properties including phonon frequencies and elastic constants[29].

Consequently, the functions generated by each method are quite different even though they are all designed to model the same element, and all yield similar values for the cohesive energy, bulk modulus, and lattice parameter of FCC gold.

The differences in the EAM functions produced are especially visible in the pair interaction components of each potential, shown in Figure 2a. While the potentials from Zhou and Grochola[5,27] are slightly negative near the FCC equilibrium distance (≈3 Å), the Ackland potential pair term is positive below 3.25 Å and the Sheng potential has a global minimum at 1.5 Å. In order to adjust each potential into a standardized form, an invariant transformation was applied (see section 2.2.1) so that the pair function has a minimum equal to 4% of the cohesive energy. This value was chosen to be consistent with other potentials in literature with Morse-like pair functions, such as those by Zhou[5], which were found to have minima between 3-5% of the cohesive energy. After applying the standardization transformations, the pair functions from each potential (shown in Figure 2b) are qualitatively similar to a Morse function, with strong repulsion at short distances and weak attraction at higher distances. Assuming that an optimal cross-potential pair function will have a shape similar to the pair functions from the pure element potentials, this transformation should increase the accuracy of the binary alloy potential.

Depending on the fraction of the cohesive energy chosen for the minimum of the pair function, the magnitudes of the fitness for each potential were found to change. This is unsurprising given that the transformations used to standardize potentials were originally used as fitting parameters when creating an alloy potential[7]. However, it would be impractical to use the standardization transformations as fitting tools as it would alter the elemental components differently in each binary potential, thereby preventing the generation of higher-order potentials from a binary database.

The optimized fitness parameter of each binary potential, calculated using Equation 13, created from each elemental potential in the original and standardized form is shown in Figure 3. Potential standardization is a key feature of this method as it allows potentials from different authors to be used effectively regardless of their choice of functional form. This is evidenced by the fact that the fitness parameter of the alloy potentials created with a gold potential from Ackland increased from 2.2 to 5.2 after applying the transformation, which makes it competitive as a selection for the gold component of the Cu-Au potential. It was also found that using potentials from the same author for copper and gold yield the best alloy potentials, which is presumably due to the choice of the same functional form for the embedding and election density functions. Even so, the other Cu-Au binary potentials created using the Zhou, Ackland, or Grochola gold potentials have competitively high finesses after the transformation.

This further demonstrates that our standardization method considerably lessens the penalty of using potentials from different authors.

It was found that not all elemental potentials are suitable for creating alloy potentials using our method. For instance, the binary potential created using a gold potential developed by Sheng *et al.* has a notably lower fitness that is worsened by standardization. The incompatibility can be attributed to the strongly different functional form used for the pair potential, with the minimum in the pair function originally two orders of magnitude lower than any other tested gold potential. While it would be impractical to prescribe a standard shape to produce optimal alloy potentials based on our work, it is possible to use our potential fitting method as a simple check to test the feasibility of a new elemental potential for use in generating a library of alloy potentials.

### 3.3. Comparison with other Alloy Potentials

To benchmark our proposed method against current literature standards, we have developed a potential describing the Cu-Zr-Al ternary system using elemental potentials from a database developed by Zhou *et al.* [5]. This potential will be compared with the ternary potential generated using identical pure potentials combined using the Johnson alloy model, as originally proposed by the author[5]. It will also be compared with a Cu-Zr-Al potential developed by Sheng *et al.* with all functions fit simultaneously against an extensive library of crystalline and amorphous structures using the force-matching method[30,31].

#### *3.3.1.* Accuracy in Modeling Intermetallics

In order to evaluate each potential's effectiveness in modeling intermetallics, the properties of all possible $L1_0$, $L1_2$, $L2_1$, B1, and B2 structures in the Cu-Zr-Al ternary were first calculated using corrected DFT, as described earlier. In total, the entire testing set contains 9 intermetallics used for fitting our potential (3 B2 and 6 $L1_2$) and 9 others, including 3 ternary compounds. The properties of those intermetallics were then calculated using all three potentials, with the median errors shown in Figure 4. A complete comparison of the properties calculated by each potential is available in Table 2.

Our method was found to be comparable to the Sheng potential and far superior to the Johnson alloy model in replicating the mixing enthalpy and bulk moduli. When taking all 18 of the tested intermetallics into account, our method shows a median error of 0.057 eV in mixing enthalpy which is comparable to 0.077 eV obtained by using Sheng *et al.* potential and far below the median error of 0.385 eV of the potentials created by Zhou. Additionally, our method gives a median error of 2.9 GPa for bulk modulus of the tested intermetallics, which is a significant improvement over the median error of 13.5 GPa and 16.8 GPa obtained by Sheng potential and Johnson alloy model potential, respectively. The lattice parameters predicted by all three potentials are comparably accurate with the median error being less than 2% using all potentials. It is worth noting that the Sheng potential was found to have the smallest maximum error, with a maximum error of 5.7% for the $a$-direction lattice parameter of $L1_0$ AlZr. In comparison, the potentials created in this work and with the Johnson alloy model have a maximum error in this lattice parameter of 14.6% and 15.0%, respectively. So, it is evident that creating a sophisticated potential with a large training set will diminish the chances of having outliers by ensuring more compounds and bonding environments are sampled and fitted against. However, taking all intermetallics into account, our potentials demonstrate predictive capability comparable to a more sophisticated and computationally expensive potential and far superior to potentials developed using the Johnson alloy model.

### 3.3.2. Effectiveness in Simulating Metallic Glasses

In recent years, molecular dynamics has seen increasing use for the simulation of metallic glasses. In fact, many potentials are specifically tuned to accurately replicate the properties of liquids and amorphous alloys[32,33]. In order to validate our method in reference to this application, a model of a $Cu_{45}Zr_{45}Al_{10}$ metallic glass was generated with each of the candidate potentials.

Each model was generated starting from the same randomly-seeded BCC lattice. The lattice was thermalized at 300 K, and then rapidly heated to 2000 K at $2.83 \times 10^{13}$ K/s. The system was equilibrated for 60 ps at that temperature and then quenched to 1500 K at $6.67 \times 10^{11}$ K/s. The model was then cooled to 200 K at an average rate of $1.0 \times 10^{11}$ K/s in temperature steps of 32.5 K. Starting at 1272.5 K, the quench rate was slowed by 12% until 622.5 K, which is well below the calculated glass transition temperature near 740 K. After this point, the quench rate was returned to the pre-1272.5 K value. This non-linear quench schedule allows for minimal simulation time to be spent where there is either quick structural relaxation or frozen dynamics, while a slow quench rate is used around the glass-transition temperature, where it is crucial[34]. The final structure at 200 K was then relaxed to 0 K using a conjugate-gradient minimization technique. The elastic constants of the 0 K structure were determined from the stress resulting from strains ranging from -0.8% to 0.8% in steps of 0.4% along all three axes.

As a first test, the 0 K structure predicted by each potential was studied to determine if it was amorphous and to determine if any anomalous behavior occurred. All three potentials produced an amorphous structure, as evidenced by the non-zero minimum between the first and second peaks of the radial distribution function in Figure 5. However, as shown in Figure 6, the Johnson alloy model predicts demixing of aluminum, which was found to occur immediately after melting. Even if demixing of the liquid was to occur at high temperature, such rapid and complete separation (which occurred in less than 20 ps at 2000 K) is unrealistic, especially considering that this composition is known for forming metallic glasses[35]. This anomaly is most likely attributable to the large discrepancies in the predicted mixing enthalpies from the Johnson alloy model. While this does not prove that the potentials developed using the Johnson alloy model are ineffective in modeling glasses in general, it does show it to be ineffective in modeling the Cu-Zr-Al ternary and suggests that its reliability may be limited overall.

The second test was to compare the elastic moduli and density predicted for the metallic glass with experimental values, as shown in Table 3. In terms of density, the potential developed by Zhou replicates the experimental value most accurately with only a 0.8% difference even though the predicted demixed structure is unrealistic. In contrast, the other two potentials predict densities between 4% and 5% lower than experiment, which is more plausible than the Johnson potential given that the density of an amorphous phase decreases with increasing quench rate. Additionally, the elastic moduli predicted by each potential are lower than experimental values, which can also be attributed to the extreme quench rate[36]. Even though there is a difference of 10% between the predicted elastic moduli of the different potentials, the fact that there is no experimental data from metallic glasses quenched at such high rates makes it imprudent to discredit any potential on the basis of the predicted properties. The relative agreement of our potential with the Sheng potential demonstrates that our method can be used to generate models of metallic glasses that replicate properties just as well as those created by more sophisticated techniques. And, given the lack of anomalous demixing, we also assert that our method for generating alloy potentials is superior to mixing rules when modeling metallic glasses.

4. Advantages and Limitations of the Method

One of the main limitations of this method is the small training set used to develop the binary potentials. More thorough studies, such as Sheng's potentials developed in[30], use a greater number of intermetallic compounds and other structures in their training sets. The broader range of atomic configurations evaluated by the fitting program directly corresponds to a denser set of pair distances and electron density values used in fitting the pair and many-body components of the EAM potential. Our method, in its current form, is not intended to generate "best-in-class" potentials but instead to enable the rapid evaluation of the alloy systems by reducing the size of the training set. However, we have found that the quality of our potentials for the test cases is very high – significantly better than the Johnson alloy potentials and very comparable to the "best-in-class" potentials. The small number of *ab initio* calculations required for a single binary system can be comfortably performed on a multi-core workstation, and are trivial on even a modest high-performance computer. Using that data to generate a binary potential only requires a few hours on a workstation and combining the binaries completes in seconds.

In addition to the reduced training set, our method is also limited by a dependence on the original potentials. This includes varying degrees of compatibility between elemental potentials developed by different authors, which has been address by using standardization transformations. Additionally, our potentials are limited by any inaccuracies present in the elemental potentials. This problem is present in any alloy potential that did not have all components fitted simultaneously, and simply requires the selection of elemental potentials suitable to replicate properties of interest in pure systems.

One key goal of this work was to make the developed alloy potential database and the computational tools freely available and easily accessible. A dedicated website has been developed to provide access to our database and for downloading our tools[37]. The Rapid Alloy Method for Producing Accurate, General Empirical (RAMPAGE) Potential Making Toolkit comes packaged with all of the software required to quickly implement a new alloy potential database if, for instance, more intensive fitting is required or a specific problem. All components used in creating the fitting tool are written in languages with free compilers or interpreters (Python, C/C++, and Octave) and utilize LAMMPS, an open-source molecular dynamics package[38,39]. The *ab initio* calculations were performed with VASP[40], though open-source options do exist[41].

5. Conclusions

In this work, we present a method for rapidly creating alloy potentials with minimal computational costs. The alloy potentials are created by using the pure-element interaction terms from previously developed potentials and fitting the binary interactions in the Finnis-Sinclair formulation of the EAM to *ab initio* data. These binary potentials can be combined to form multi-component potentials that maintain the accuracy of the original elemental and binary systems with no additional fitting. Potentials created using our method were demonstrated to be superior to alloy potentials generated using the Johnson alloy model and comparable in accuracy to potentials that fit all interaction terms simultaneously against a larger training set. A library of alloy potentials for a large number of metallic elements has been made available on the internet[37].

6. Acknowledgements

This work was funded by the Defense Threat Reduction Agency under Grant No. HDTRA1-11-0047 and by the Air Force Office of Scientific Research under Grant No. FA9550-09-1-0251. Simulations were performed at the Ohio Supercomputer Center (Grant No PAS0072). LW was supported by the Department of Defense (DoD) through the National Defense Science & Engineering Graduate Fellowship (NDSEG) Program.

7. Figures

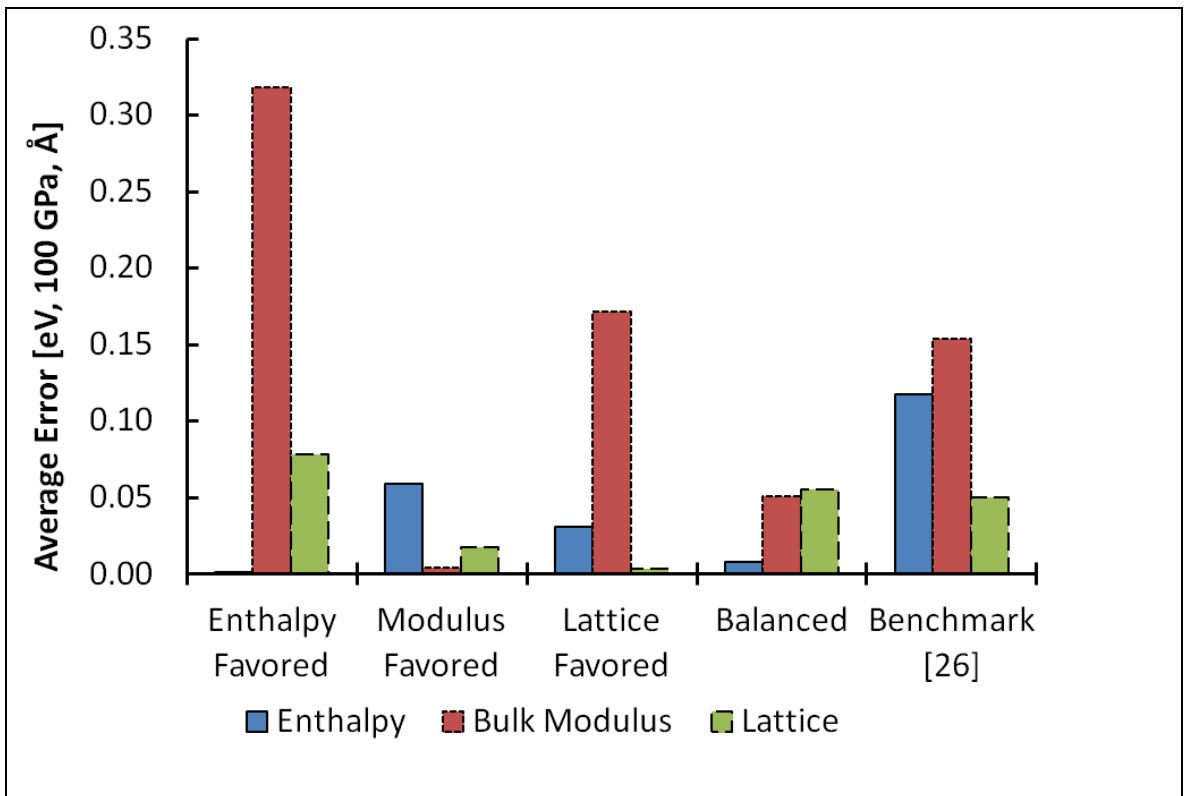

**Figure 1.** Average deviation between properties for B2-AlNi, L1$_2$-Al$_3$Ni, and L1$_2$-Ni$_3$Al intermetallics calculated using corrected *ab initio* and EAM potentials created with several weighting strategies. The average difference between an Al-Ni potential by Pun and Mishin and corrected *ab initio* is shown for comparison[20].

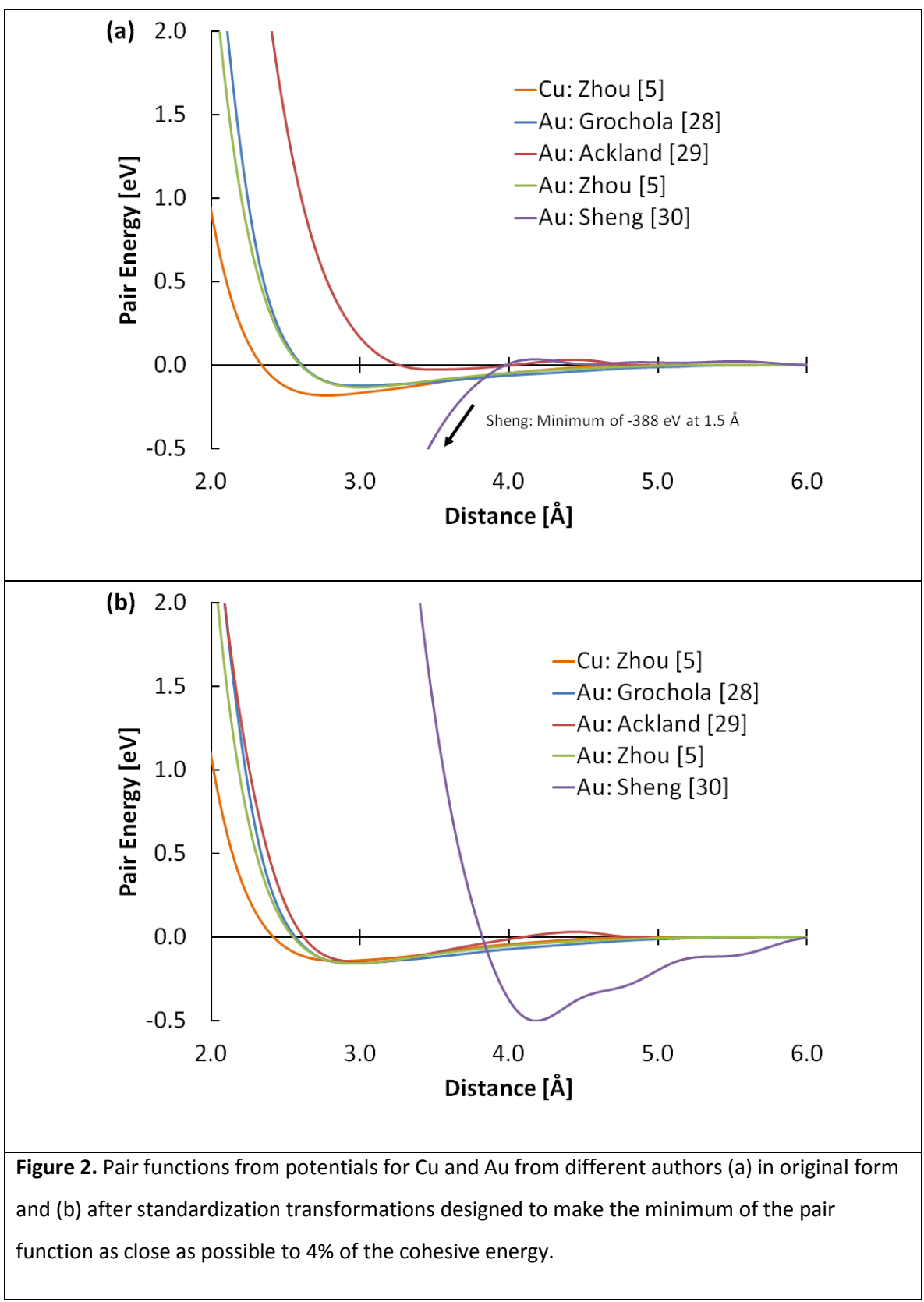

**Figure 2.** Pair functions from potentials for Cu and Au from different authors (a) in original form and (b) after standardization transformations designed to make the minimum of the pair function as close as possible to 4% of the cohesive energy.

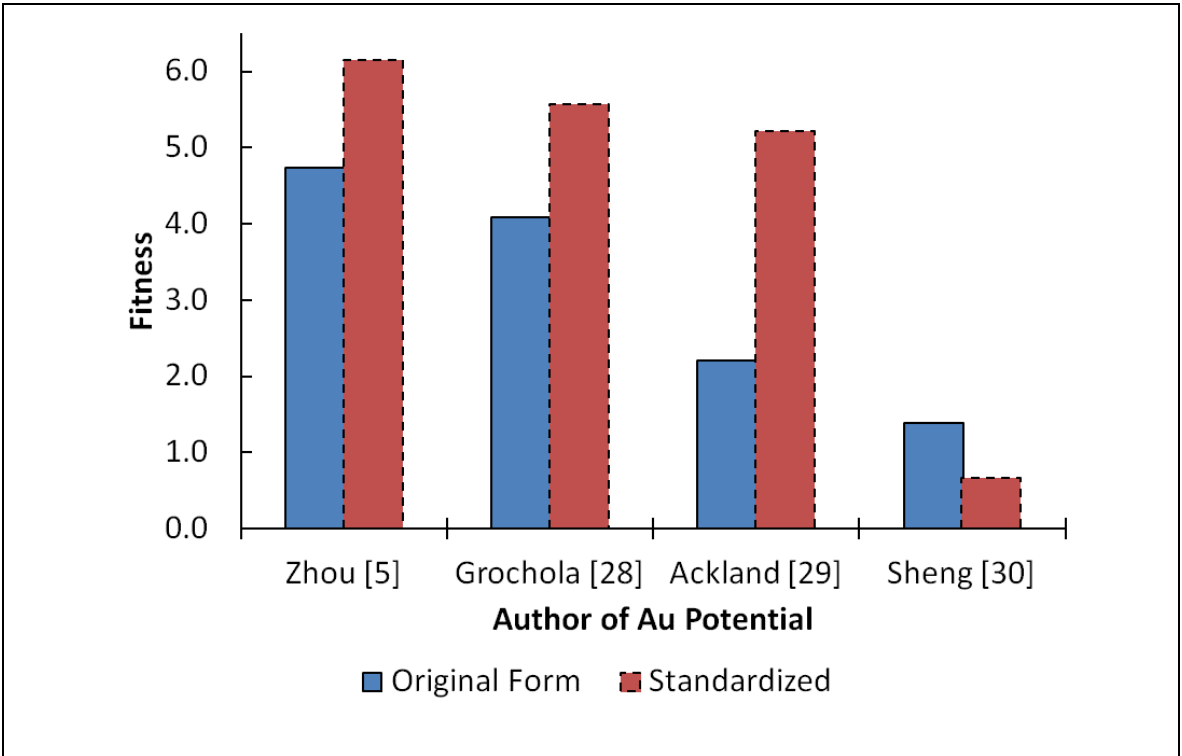

**Figure 3.** Overall fitness of Cu-Au EAM potential generated in the present work with and without standardization transformations made with a Cu potential from Zhou *et al.*[5] and Au potentials from different authors, as indicated.

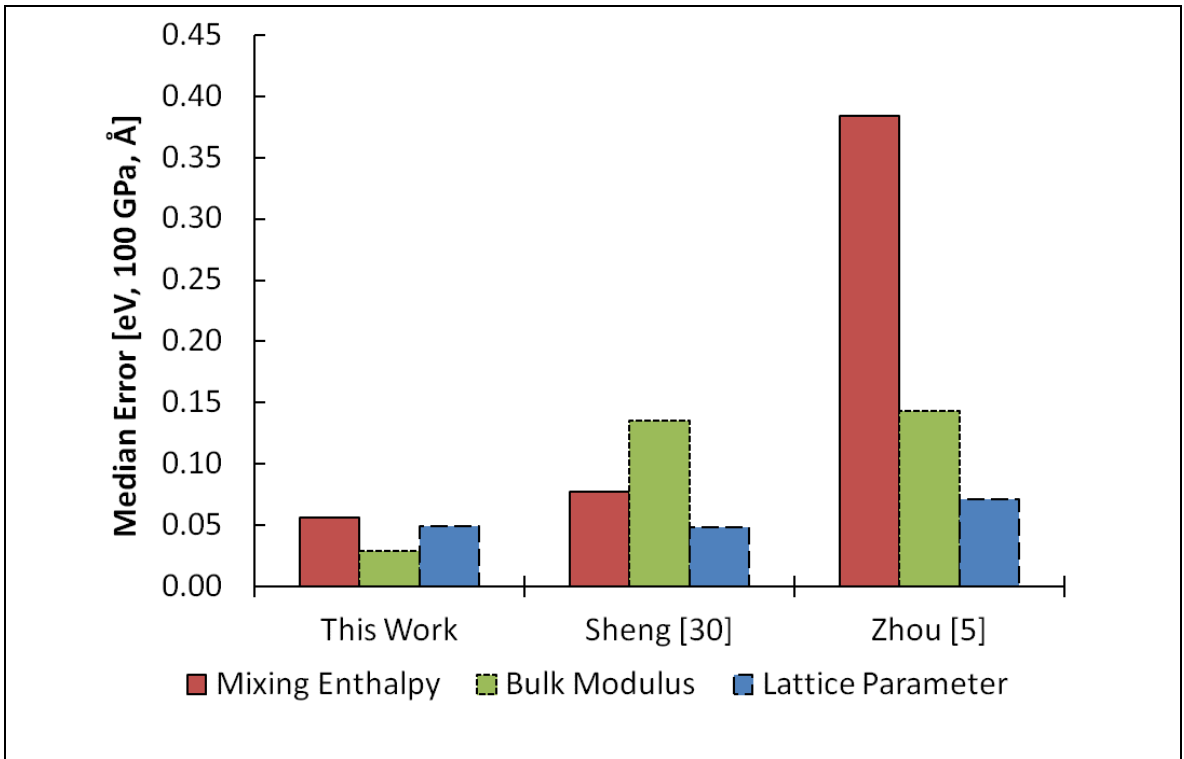

**Figure 4.** Median error in mixing enthalpy, bulk modulus, and lattice parameter in reference to *ab initio* data for Cu-Zr-Al potentials with alloy components fitted in this work, developed by Sheng [30], and potentials created using the Johnson alloy model [5]. The intermetallics used in testing include all 18 possible B1, B2, $L1_0$, $L1_2$, and $L2_1$ compounds.

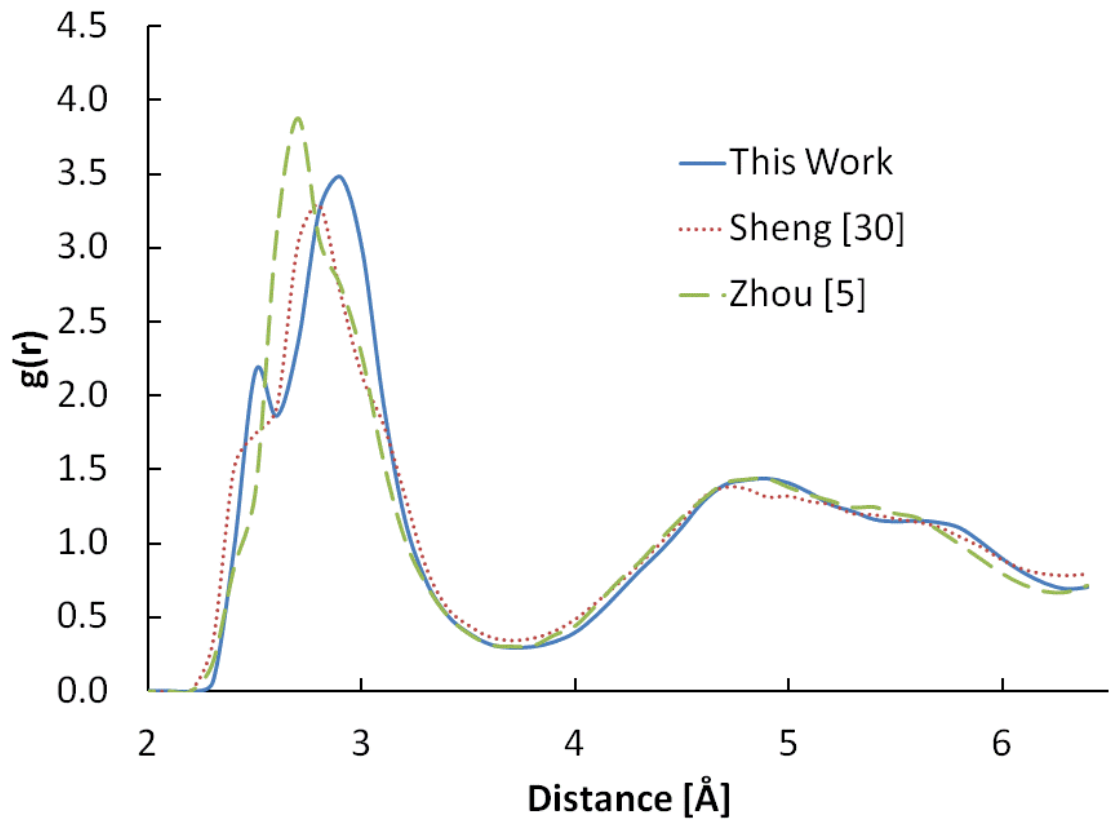

**Figure 5.** Radial distribution functions of $Cu_{45}Zr_{45}Al_{10}$ metallic glasses at 0 K produced using interatomic potentials developed in this work, by Sheng[30], and Zhou using the Johnson alloy model[5].

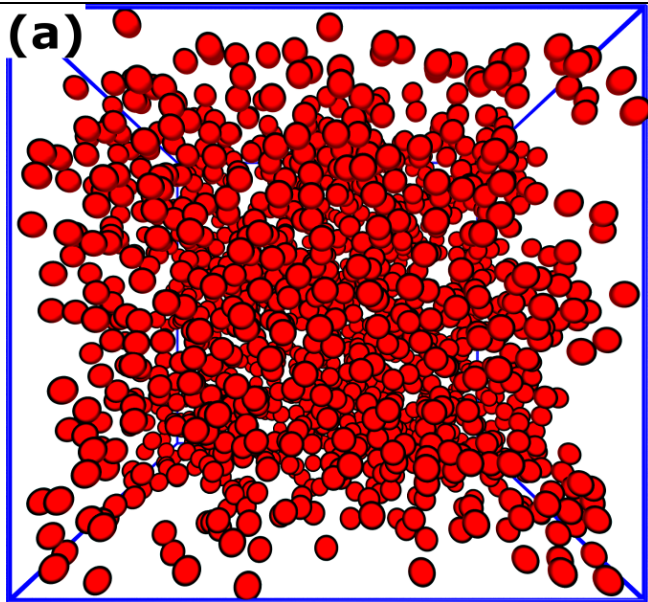

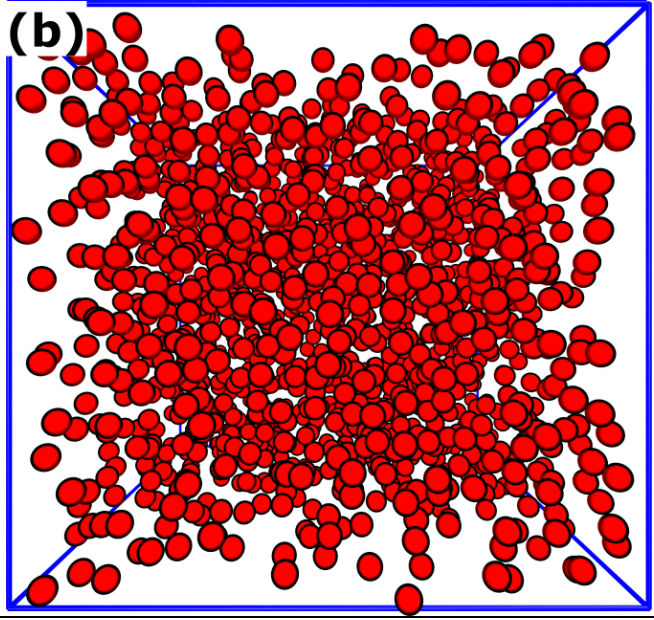

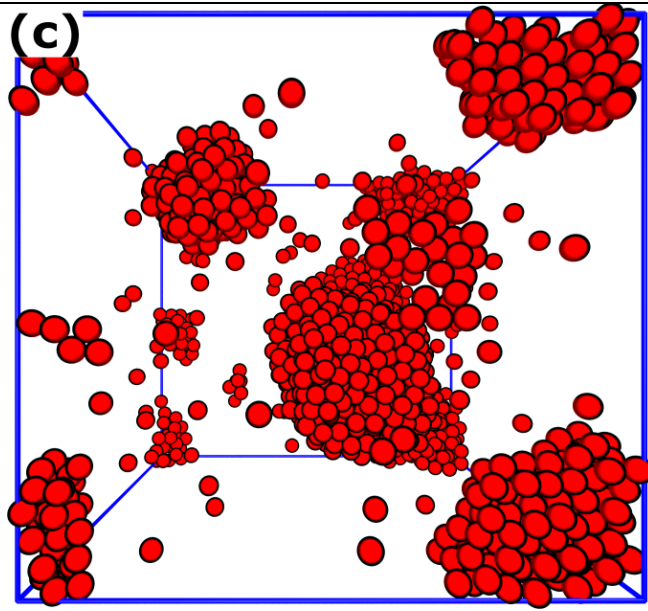

**Figure 6.** Aluminum distribution in $Cu_{45}Zr_{45}Al_{10}$ metallic glass at 0 K in 16000-atom models generated by simulating a rapid quench from the liquid state using potentials made by (a) this work, (b) Sheng[30], and (c) the Johnson alloy model[5]. This potential created in our work and by Sheng show a uniform distribution of aluminum. In contrast, the Johnson alloy potential predicts a distinct demixing the aluminum, which is not likely for this alloy composition[35].

8. Tables

**Table 1.** Comparison of mixing enthalpy, bulk modulus, and lattice constant for fcc-Ni, $L1_2$-$Ni_3Al$, B2-NiAl, $L1_2$-$Al_3Ni$, and fcc-Al in the Ni-Al system; and fcc-Cu, $L1_2$-$Cu_3Au$, B2-CuAu, $L1_2$-$Au_3Cu$, and fcc-Au, calculated using DFT within GGA and EAM potentials fitted in the current work in comparison with experimental data.

| Compound | Mixing Enthalpy [eV/atom] | | | Bulk Modulus [GPa] | | | | Lattice Constant [Å] | | | |
|---|---|---|---|---|---|---|---|---|---|---|---|
| | EAM | DFT | Experiment | EAM | DFT | DFT, Corrected | Experiment | EAM | DFT | DFT, Corrected | Experiment |
| fcc-Ni | - | - | - | 178. | 202. | 186. | 186.[f] | 3.52 | 3.52 | 3.52 | 3.52[f] |
| $L1_2$-$Ni_3Al$ | -0.445 | -0.444 | -0.387[a] | 172. | 183. | 173. | 175.[c] | 3.51 | 3.57 | 3.58 | 3.57[e] |
| B2-NiAl | -0.667 | -0.675 | -0.609[a] | 167. | 162. | 157. | 166.[d] | 2.85 | 2.89 | 2.90 | 2.88[e] |
| $L1_2$-$Al_3Ni$ | -0.238 | -0.223 | - | 106. | 111. | 110. | - | 3.80 | 3.85 | 3.85 | - |
| fcc-Al | - | - | - | 72.7 | 74.6 | 75.9 | 75.9[f] | 4.05 | 4.05 | 4.05 | 4.05[f] |
| fcc-Cu | - | - | - | 135. | 143. | 138. | 138.[f] | 3.62 | 3.63 | 3.62 | 3.62[f] |
| $L1_2$-$Cu_3Au$ | -0.042 | -0.042 | -0.070[a] | 141. | 143. | 148. | 152.[c] | 3.74 | 3.79 | 3.76 | 3.75[e] |
| B2-CuAu | -0.040 | -0.049 | - | 157. | 138. | 153. | - | 3.04 | 3.12 | 3.08 | - |
| $L1_2$-$Au_3Cu$ | -0.023 | -0.022 | -0.060[b] | 163. | 140. | 164. | 166.[c] | 3.97 | 4.06 | 3.99 | 3.97[e] |
| fcc-Au | - | - | - | 166. | 139. | 173. | 173.[f] | 4.08 | 4.17 | 4.08 | 4.08[f] |

[a] Ref 21, values at ambient temperature
[b] Ref 22 for compound data, 23 for elemental
[c] Ref 24
[d] Ref 25
[e] Ref 26
[f] Ref 16

**Table 2.** Mixing enthalpy, bulk modulus, and lattice parameter for all 18 possible B1, B2, L1$_0$, L1$_2$, and L2$_1$ intermetallics in the Cu-Zr-Al system calculated using DFT and potentials developed in this work, by Sheng[30], and Zhou using the Johnson alloy model[5]. The potentials developed in this work were created by fitting to the DFT properties from the B2 and L1$_2$ intermetallics.

|  | Compound | Mixing Enthalpy [eV] | | | | Bulk Modulus [GPa] | | | | Lattice Parameter [Å] | | | |
|---|---|---|---|---|---|---|---|---|---|---|---|---|---|
|  |  | DFT | This Work | Sheng[30] | Zhou[5] | DFT | This Work | Sheng[30] | Zhou[5] | DFT | This Work | Sheng[30] | Zhou[5] |
| Training Set | B2-AlCu | -0.139 | -0.140 | -0.293 | -0.076 | 111.2 | 111.6 | 136.5 | 69.4 | 2.992 | 2.956 | 2.941 | 3.036 |
|  | B2-AlZr | -0.318 | -0.449 | -0.397 | 0.067 | 104.9 | 104.1 | 98.5 | 106.0 | 3.405 | 3.401 | 3.356 | 3.455 |
|  | B2-CuZr | -0.138 | -0.043 | -0.137 | -0.778 | 119.8 | 120.3 | 102.9 | 109.0 | 3.283 | 3.279 | 3.253 | 3.190 |
|  | L1$_2$-Al$_3$Cu | -0.040 | -0.043 | -0.111 | -0.086 | 92.4 | 92.6 | 91.8 | 85.1 | 3.934 | 3.898 | 3.859 | 3.963 |
|  | L1$_2$-Al$_3$Zr | -0.475 | -0.446 | -0.348 | 0.085 | 104.2 | 92.3 | 99.1 | 64.2 | 4.121 | 4.119 | 4.128 | 4.192 |
|  | L1$_2$-Cu$_3$Al | -0.177 | -0.101 | -0.237 | -0.105 | 128.8 | 128.9 | 150.3 | 105.0 | 3.683 | 3.660 | 3.645 | 3.757 |
|  | L1$_2$-Cu$_3$Zr | 0.110 | 0.107 | -0.023 | -0.569 | 116.4 | 112.0 | 114.8 | 108.6 | 3.943 | 3.931 | 3.895 | 3.797 |
|  | L1$_2$-Zr$_3$Al | -0.342 | -0.332 | -0.265 | 0.042 | 105.2 | 89.7 | 108.9 | 84.8 | 4.412 | 4.461 | 4.370 | 4.474 |
|  | L1$_2$-Zr$_3$Cu | -0.012 | -0.012 | 0.056 | -0.416 | 101.5 | 102.3 | 83.5 | 93.8 | 4.340 | 4.340 | 4.328 | 4.311 |
| Validation Set | B1-AlCu | 0.193 | 0.229 | -0.004 | 0.180 | 78.9 | 101.3 | 101.8 | 86.6 | 5.049 | 4.909 | 4.997 | 5.068 |
|  | B1-AlZr | -0.023 | -0.449 | -0.066 | 0.433 | 85.7 | 82.3 | 106.2 | 101.9 | 5.603 | 5.449 | 5.579 | 5.612 |
|  | B1-CuZr | 0.228 | 0.504 | 0.150 | -0.297 | 106.1 | 114.5 | 137.1 | 123.6 | 5.391 | 5.368 | 5.308 | 5.190 |
|  | L1$_0$-AlCu | -0.147 | -0.139 | -0.294 | -0.133 | 110.3 | 111.2 | 133.3 | 104.6 | a=4.107, c=3.348 | a=4.180, c=2.956 | a=4.010, c=3.181 | a=3.934, c=3.850 |
|  | L1$_0$-AlZr | -0.469 | -0.586 | -0.429 | 0.067 | 107.4 | 99.8 | 109.0 | 86.0 | a=4.198, c=4.328 | a=4.118, c=4.959 | a=3.957, c=3.392 | a=4.383, c=4.369 |
|  | L1$_0$-CuZr | -0.138 | -0.038 | -0.137 | -0.778 | 121.4 | 127.6 | 102.9 | 109.0 | a=3.282, c=4.643 | a=3.578, c=4.439 | a=3.252, c=4.600 | a=3.190, c=4.512 |
|  | L2$_1$-Cu$_2$ZrAl | -0.369 | -0.171 | -0.283 | -0.380 | 128.7 | 99.5 | 121.6 | 93.4 | 6.263 | 6.303 | 6.202 | 6.247 |
|  | L2$_1$-CuZr$_2$Al | -0.288 | -0.266 | -0.252 | -0.338 | 114.3 | 114.6 | 109.0 | 108.4 | 6.701 | 6.649 | 6.607 | 6.673 |
|  | L2$_1$-CuZrAl$_2$ | -0.117 | -0.381 | -0.221 | -0.100 | 97.4 | 99.7 | 107.5 | 77.7 | 6.442 | 6.281 | 6.382 | 6.560 |

**Table 3:** Elastic and physical properties of a $Cu_{45}Zr_{45}Al_{10}$ metallic glass predicted using molecular dynamics in comparison to experimental data[35]. The potentials include one generated in this work, a potential holistically fitted to alloy properties and the force-matching approach[30], and a potential developed within the Johnson alloy model[5].

|  | Elastic Moduli [GPa] | | Density at 300 K [g/cc] |
| --- | --- | --- | --- |
|  | Elastic | Shear |  |
| Sheng[30] | 72.0 | 25.8 | 6.93 |
| Zhou[5] | 64.4 | 23.1 | 7.15 |
| This Work | 66.2 | 23.7 | 6.81 |
| Experiment[35] | 99.1 | 36.3 | 7.20 |